# A Survey of Bandwidth Optimization Techniques and Patterns in VoIP Services and Applications

Uchenna Peter Daniel[1], Nneka Chikazo Agbanusi[2] and Kwetishe Joro Danjuma[3]

[1]Department of Computer Science, Federal University Lokoja,
PMB 1154, Kogi State. Nigeria

[2]H. Pierson Associates Ltd, Lagos -Nigeria

[3]Department of Computer Science, Modibbo Adama University of Technology Yola
Adamawa State, Nigeria

**Abstract**
As Internet Telephony unveils a communication system that allows for the conduction and reception of voice signals and data over the internet network, the volume of voice traffic and (or) packets on IP networks have grown substantially. However, a crucial problem imminent is that of inefficient bandwidth utilization which is mostly as a result of header overheads resulting from the attachment of a big header size (40 bytes) to small payload size (10 to 30 bytes). Now the broadband telephone industries reveal a lot of business prospects in revenue, with great benefits and rewards for current and could be investors. However, these service providers must continue to make out ways of keeping themselves relevant and strong amidst great competition. The quest is to condense the amount of bandwidth being wasted, or better still; improve the utilization of bandwidth which eventually enhances network performance and VoIP quality of service. Numerous schemes and techniques have been suggested to meet this need. This article surveys the various techniques adopted for optimising bandwidth for VoIP services over the period 1999-2014. The improvement of bandwidth can be realized through; silence suppression measure of repressing the silent portions (packets) in a voice conversation using Voice Activity Detection algorithm; by so doing, the transmission rate during the inactive periods of speech is reduced, and thus, the mean transmission rate can be reduced. A second measure is packet header reduction which defines a process of multiplexing and de-multiplexing packet headers to curb excesses. Voice/ Packet Header compression is considered the most productive of all the techniques, offering a scheme where VoIP packets are compressed from the 40 bytes of size to a smaller byte size of 2 bytes. When combined with aggregation, compression potentially yields a compressed size of up to 1 byte. In either case, bandwidth save is reached using compression and decompression codecs of varying data and bit rates. It is envisaged that an improvement in the performance of codecs would yield a better result in terms of enhancing results favourably in Voice over broadband networks.

## I. INTRODUCTION

The volume of voice traffic on IP networks continues to increase at speedy strides, with substantial growth in the use Skype, Tango and other voice applications. Voice over Internet Protocol (VoIP) also called also called, IP Telephony, Internet telephony, Broadband telephony, Broadband Phone and Voice over Broadband [1] a technology that unveils a communication system that allows for the conduction and reception of voice signals and data over the internet network. While using the Internet Protocol as the most basic transport mode over which both TCP and UDP are utilized [2], VoIP system uses designated codecs that convert voice signals into digital data forms (bits) [3] yielding an output that is transmitted through networked infrastructure over the internet. These bits are usually reconstructed at the destination using attribute data and timestamps accordingly.

Equitable cost savings, ease of deployment while leveraging on existing infrastructure [4], scalability, improved productivity, flexibility and mobility [1] are key drivers for the proliferation of this technology. FCC predictions were that by 2008, 44% of corporate organizations were going to run VoIP lines [4].

It is so envisioned that in time, VoIP technology (applications and schemes) will substitute traditional Public Switched Telephone Network (PSTN) Technology [5] in [6]. This ascendency over PSTN is and will continue to be driven by numerous factors; notably higher reliability due to systematic and automated side-stepping of inherent challenges of network over congestions; with the ability to execute calls, regardless of geographical location using computers or other portable mobile devices [7].

However, voice traffic operates in real-time and is centered on small UDP packets, thus, incredible network loads are exerted attendant devices due to the emergence of huge packets per second per voice call [8]. Additively, voice applications that are VoIP-enabled are greatly inefficient in bandwidth given that packet header sizes are often same as the payload size with overheads of nearly 40% [8].

More so, acceptable use standards advocate for efficient Quality of Service, a paradigm that plainly describes the gratification level experienced by users of any VoIP application. Technically, a user will recognize fragmentations in voice transmission and poor quality when high delay, loss or distortions occur. Therefore, choices for packet sizes are made based on network delays, through accessible bandwidth and packet losses inside the network. One very crucial problem in a VoIP network is the potential wastage of bandwidth caused by the earlier noted inefficiency also caused by attachment of huge (typically 40 bytes) unwanted bandwidth to smaller payloads ( typically 10 to 30 bytes) desirable bandwidth (payloads). Consequently, the need to reduce bandwidth wastage; or better still improve the utilization of bandwidth which could eventually boost VoIP quality and network performance cannot be overstated.

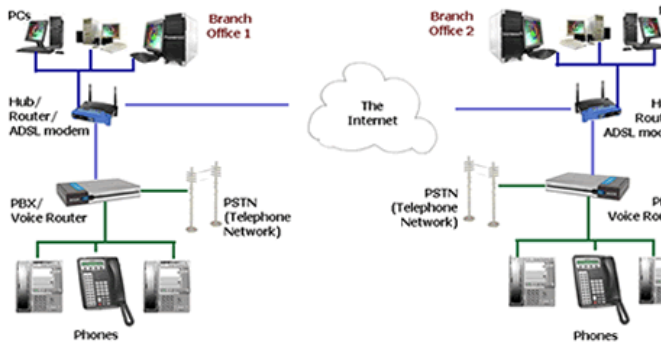

**Figure 1: VoIP Setup [1]**

Maintaining excellent quality voice services to users is a major concern for VoIP service providers [9] in [10]; a VoIP service that is fast (no or minimal delay) and efficient. However, "accelerating" VoIP traffic is quite vague, and often misleading. Why? Because VoIP traffic like all other IP traffics, moves at the speed of light. It can thus be delayed by network congestion and other relative issues that could amount to despoiled voice quality. Nonetheless "accelerating" VoIP traffic makes no contextual sense since VoIP packets cannot be made to travel faster than the current speed of light, and (or) participants cannot be made to talk any faster than they are able on telephony applications [11]. From a technical viewpoint and following tolerable standards, VoIP datagrams must be conveyed by the network with a negligible amount of jitter in order to preserve its quality. Latency must be retained below 150ms [11] if possible, and any techniques adopted to expedite traffic delivery must eliminate or reduce latency and jitter to the nearest minimum. Nonconformities to these standards will explicitly affect voice quality. This paper survey the various schemes ad techniques adoptable for the optimization of bandwidth in Voice-Over Internet Protocol Networks and (or) Services

Several schemes and(or) techniques have been suggested as viable solutions for enhancing bandwidth usage, and this discourse seeks to review the varying techniques with a view to ascertaining respective weaknesses. Figure 1 illustrates a sample VoIP Setup while figure 2 shows the VoIP protocol stack with respect to its TCP/IP protocol standard [12] in [13]

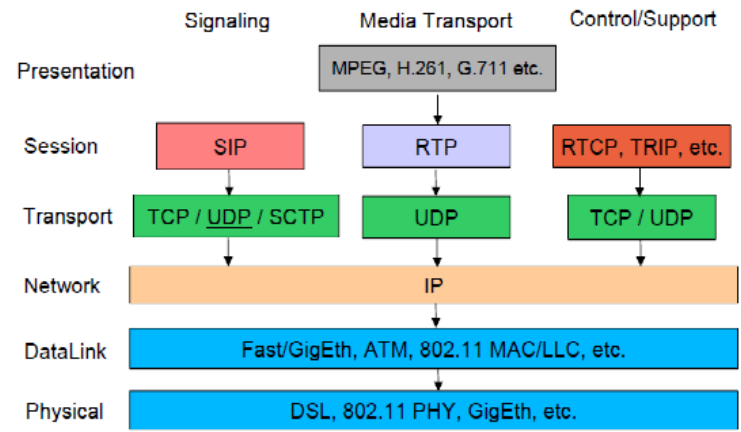

**Figure 2: VoIP implementation in TCP/IP protocol standard [12] in [13]**

## II. VOIP BANDWIDTH OPTIMIZATION: SPURS AND TECHNIQUES.

### A. Why Optimization?

Bandwidth optimization, especially with respect to VoIP systems is a crucial need. Confronted by an extremely competitive data and VoIP industry phenomenon, as well as the realities of weakening revenues for voice services per user, communication operators and service providers are looking for solutions that can lessen the odds while at the same time putting them in a stronger, success-enabling position [14]**.** There are so much reasons and ways from which these companies need and can benefit from optimization. First, it is that nearly 40% of most VoIP packets are headers. Even so, most headers hardly change, causing redundancies during transmission. Now, these redundancies continue with call frequencies despite similarities in source and (or) destination IPs, with no provision for sharing [8].

Other reasons include the fact that Ineffectiveness applies to all UDP streams, from standard VoIP applications like Skype to smaller packet application that do not use full frame video streaming. Evidently, these real-time streams require management and prioritization for effective quality-size tuning, for minimal impact and quality improvement [8].

*B. Optimization Techniques*

There are essentially three major techniques for improving bandwidth utilization in a VoIP network/service, which include; header compression/suppression (sometimes with coalescing, aggregation or prioritization), packet header reduction, and Silence suppression.

*1) Voice / Packet Header Compression*

This refers to the phenomenal representation in which voice packets are compacted from the default 40 bytes size to a smaller 2 bytes size in best. Compression and decompression codecs are used to achieve this. Codecs convert the voice signals from analog to digital, and additionally compress the resultant digital voice using digital compression algorithms. The compressed information is transformed to frames (packet payload) of varying sizes depending on the type of codec being used and its efficiency [15].

Header compression for bandwidth optimization is motivated by some basic actualities. First, VoIP datagram is typically compressed at the application layer, implying that no other compression is required to condense payload size. Secondly, a huge portion of the data packet size is occupied by the headers, thirdly, significant redundancies usually exist in IP headers which could be best managed via compression means with recourse to efficient algorithms and(or) protocols [4]. These protocols as would be seen do help reduce greatly the size of data, and there are numerous compression codecs with varying compression rates as could be seen in table 1 below.

**Table 1: Voice Codecs Listings [15]**

| Codec | Frame Size | Algorithm Delay | Compressed Rate (Bitrate) / kbps |
|---|---|---|---|
| G.723.1 (lr) | 30 | 37.5 | 5.3 |
| G.723.1 (hr) | 30 | 37.5 | 6.3 |
| G.729 | 10 | 15 | 8 |
| G.729A | 10 | 15 | 8 |
| G.729D | 10 | 15 | 6.4 |
| G.729E | 10 | 15 | 11.8 |
| iLBC (lr) | 30 | 30 | 13.33 |
| iLBC (lr) | 20 | 20 | 15.2 |
| Speex | 20 | 30 | Various |
| GSM-FR | 20 | 20 | 5 |
| GSM-HR | 20 | 20 | 24 |
| GSM-EFR | 20 | 20 | 18 |
| AMR | 20 | 25 | Various |

The key backings for header compression is the reduction of the amount of control information in packet headers and subsequently decreasing its portion in the overall packet size, and increasing bandwidth availability for data (voice/payload) transportation. Redundant details of each flow are captured and stored in a data structure on both the source (compressor) and destination (de-compressor) [16].

Quite clear to grasp is that header compression exploits the VoIP packet field characteristics of either being unchanged or increasing at persistent ratio all through a call period. Compression/transmission protocols (CRTP, ROCCO, RoHC, etc.) are what help to effectively ensure source to destination movement of voice packets. With the help of such protocols as would be reviewed further in this work, this technique does meaningfully improve bandwidth utilization. Figure 3 shows a pictorial representation of the compression technique.

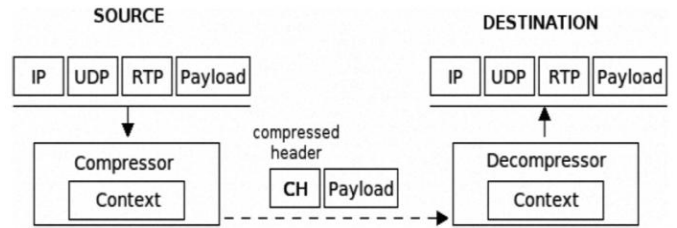

**Figure 3: Header Compression Representation [16]**

*1.1) Compression with Aggregation*

Since current header compression schemes are seen to maintain contexts with all kinds of information amounting to high level compressions, there is the need to achieve lower compression and higher speech quality, which could be attained through the utilization of algorithms that do not require recent memory yet realizing reasonable compression gains [16]. A possible tend towards such solutions is the combination of header compression scheme with packet aggregation. This offers an alternative potential for eliminating redundant information from headers. The initial compression is usually supported by a second compression by collating redundant dynamic information between headers into same aggregation packets and suppressing them into a single aggregate header [16].

This approach to optimizing VoIP bandwidth was exploited by [4] while proposing the use of a zero-length header compression (ZLH) algorithm for the elimination of redundant headers on VoIP networks. Combined with aggregation, the algorithm is able to support efficient header compression allowing for VoIP calls with as much as 75 R-score. Figure 4 shows how the compression with aggregation works.

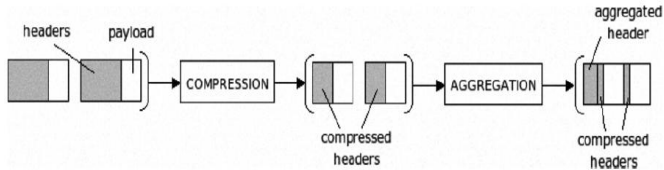

Figure 4: Compression with aggregation [16]

*1.2) Compressed Real-time Transport Protocol (cRTP) Compression*

Internet Protocol (IP) telephony services use RTP protocols for their operations, hence, IP/UDP/RTP headers of each voice packet can be compressed from the conventional 40 bytes or more to a smaller size of as low as 2 bytes with the aid of Compressed Real-time Transport Protocol, (cRTP). However effective for compression, packet loss rate at the receiver ends after each successful decompression, has been noted to be too high [17] Evidently showing that cRTP unaided might not offer the best solution to header compression in cellular & VoIP links; since packets are lost immensely [6]. A high error-rate acceptance scheme that is at least efficient as cRTP is required. Efficient headers and packet reconstruction is a much desirable feature.

*1.3) Robust Checksum-based header Compression (ROCCO) Compression*

Robust Checksum-based header Compression (ROCCO) was designed to meet cRTP's weaknesses of high packet loss at the point of decompression. ROCCO was designed to adjust favourably to the characteristics of packet stream compression and the medium through which packet loss occurs [18]. ROCCO is focused towards local de-compressor patch-up, which seeks to accomplish several reconstruction attempts to achieve the correct header. Today, ROCCO profiles for VoIP exist which are capable of compressing IP/UDP/RTP headers down to a minimum size of 1 to 2 bytes. ROCCO significantly reduces the negative effects on header compression performance that are otherwise caused by high packet loss [19]. Even more, its compression on headers is much better than cRTP and offers better security against errors on header formation [17].

*1.4) CRTP / ROCCO UDP Lite Integration.*

The UDP Lite protocol offers an elastic way for applications make assessment for possible shedding or release of packets resulting from transport to application layers bit errors. Combining this scheme with cRTP could potentially yield double the amount of packets to application when compared to classical UDP. Integrating UDP lite with ROCCO will also return a greater amount of packets to the application as simulated in [17].

*1.5) Robust Header Compression (RoHC)*

This is yet another standardized protocols defined in RFC 3095 [20] in [16] for the efficient saving of bandwidth especially in cellular network. Borne out of the need to improve on earlier protocols, Robust Header Compression came to light on the bases of integrating the working properties of three earlier protocol propositions, which include; RoHC was designed to handle bandwidth optimizations in situations of high BER and long RTT are common on 2.5G and 3G links, particularly common to 2.5G and 3G links [16].

To point out a motivation, as quoted in RFC 3095 [20], "Bandwidth is the most costly resource in cellular links. Processing power is very cheap in comparison. Implementation or computational simplicity of a header compression scheme is therefore of less importance than its compression ratio and robustness." Prior, to this, RoHC is noted to be capable of achieving packet compression up to 1 byte and thus is considered better efficient than the compression schemes that preceded it [21]. In-depth study has revealed RoHC to harbor the weakness of requiring feedback channel in two out of three context updates or sent acknowledgements [16], it is yet essential to optimize network bandwidth, reduce the packet loss due to bit errors and reduce delays due to the large overheads created. Indisputably, RoHC header compression comes crucial towards improving user experience on Voice service delivery [21].

Generally, header compression technique reveals several benefits in digital network communications. Notable rewards include; optimized networked transmission efficiency, speed and quality with reduced packet losses, decreased packet header overhead, decreased infrastructural costs while making for more users per channel and ensuring better interactive response time [22]. Relative to VoIP applications, the technique aids exceeding bandwidth savings of up to 60% [22].

*2) Packet Header Reduction*

This describes the process of reducing the size of the packet header from its usual size to a smaller size such that overall overhead is condensed for transmitted packets. This is achieved using the processes of multiplexing and de-multiplexing otherwise known as delta-multiplexing

*2.1) Multiplexing*

Multiplexing VoIP packets for payload size reduction significantly cuts overhead. [23]. The model here presents a procedure whereby the traditional 40 bytes of VoIP packet header that combines with the usual payload (10 – 30 bytes) is condensed. The combination of the huge packet header with payload usually causes huge overhead, could be technically reduced by multiplexing the related payloads in one header. By this, bandwidth is saved from wastage.

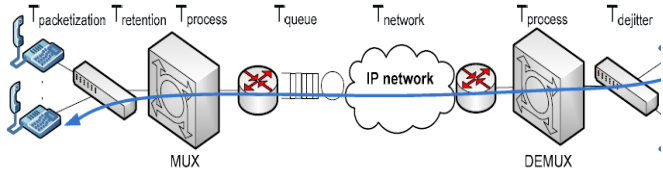

**Figure 5 : Multiplexing technique [23]**

*a) RTP Multiplexing Techniques and Applications*

Hoshi and his colleagues [24], proposed a Voice Multiplexing scheme that could be used to reduce the number of RTP packets transported over the IP Network. Their scheme combines voice packets from varying streams into a single UDP packet for transmission. Although this technique improves transmission efficiency, it does not handle the idea of compressing the RTP headers [24].

This mechanism for saving bandwidth has been leveraged upon by Sze et al [25], where the proposed a technique that explores the combination of packet multiplexing and header reduction through accumulating several RTP packets in a single UDP packet. The RTP packets are compressed for which context-mapping tables are created in the multiplexer and de-multiplexer. This is to ensure that original packets are rebuilt or restored at the destination end. Through experimentation / simulation, the proposed system showed an effective utilization of Wide Area Network trunk and an increase in the number of supported real-time calls severally, as compared to the conventional scheme [25].

A multiplexing scheme presented by [26] also expresses more light on the theory of RTP multiplexing using mixers and translators. While the mixer is tasked with collecting and adding-up of different voice streams, changing the data format and retransmitting to the receiver, the translator re-encodes multiple packets into one. In the long run, both the mixer and translator send only a combined or translated RTP flow. Thus, the TC RTP defines the combination of protocols for effective management of voice packets. ECRTP header compression scheme compresses the IP, UDP and RTP headers into one new header. Thus, final transmission of packets through respective tunnelling scheme is achieved using the PPP multiplexing.

*2.2) Delta-Multiplexing (De-Multiplexing)*

Delta-multiplexing technique is another method proposed by for improving band utilization [27]. Again we recall that bandwidth inefficiency and network overloads are two major challenging issues in a computer network circle. Delta-multiplexing provides solutions by combining header overhead reductions and payload size deductions [27].

The delta-multiplexing architecture consist of two entities; the multiplexer (Mux) which is located in the sender gateway and performs payloads size reduction and packet multiplexing. The second entity is the D-Multiplexer (D-Mux) which is located at the receiver gateway, and performs packet de-multiplexing, returning the payloads to its default size. A Performance investigation on bandwidth efficiency for multiplexing 10 users in each stream showed that a cumulative level of up to 68% - 72% of bandwidth is saved, depicting an improvement in network performance with respect to network traffic, overload and packet congestion. Running VoIP packets over network are reduced and voice quality is enhanced. This appropriately makes Delta-multiplexing compliant to SIP and H.323 systems [27].

*a) De(Multiplexing) Applications*

Multiplexing approach was applied on IP-Telephony Gateways (IP-TG) that linked PSTN/IPBX to IP networks [27]. It was also seen to be appropriate for IP-TG networks linking Cellular Access Networks (CAN) with Mobile Switching Centres (MSC). This practice multiplexes multiple VoIP packets from different sources in single RTP header. A Test application showed that resultant overhead reduced from 50% to 80%.

Another application of multiplexing was expressed in [15] involving voice packets generated by Session Initiation Protocols (SIP) applications. The scheme depended on the hypothesis that there are multiple SIP VoIP LANS linked via one SIP WAN VoIP Gateway (SWVG). This implied an increase in traffic on the SWVG gateway thus improving the multiplexing process. The concept is such that the sender SWVG gateway multiplexes the packets destined to the same destination, while the receiver SWVG gateway de-multiplexes the packets and dispatches them to their various destinations. The outcome is such that header overhead is reduced, thus saving bandwidth. And quite supplementary is that the number of packets sent is reduced influencing reduction in the total overhead on network hips [15].

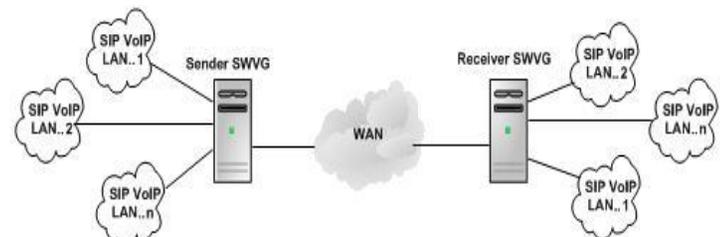

**Figure 6 : SWVG Gateway Connect Multiple SIP VoIP LAN [9]**

*3) Silence Suppression.*

This approach describes the technique for saving bandwidth wastage in a VoIP system. A silence suppression technique utilizes algorithm functions comprising of an encoder and a decoder.

On the encoder side, voice activity detection (VAD) mechanism is incorporated with the task of monitoring received signals for voice activity. If this is secured, a low bit-rate encoder is initiated before padding the transmission details. And when activity is not detected for a preconfigured time period, the encoder output is prevented from being transported across the network; resulting in additional bandwidth savings [28], [29] This is achieved using a special packet called silence insertion descriptor (SID) packet that holds some characteristic parameters of the background noise, created and released to the far end. And discontinuous transmission (DTX) algorithm defines SID packet transmission frequency. On the decoder side, a signal representative of the silence in between conversation is generated by a comfort noise generation (CNG) engine. This is meant to fill in the breaks that the original silence should have occupied. At such, the transmission rate during the inactive periods of speech is reduced, and thus, the average transmission rate is condensed as well [30]. Quite assuredly, VoIP frames will not be generated continuously for each user, since there are silent periods within talk periods. These periods are noted to follow exponential distribution such that each sound alternates between OFF-state and ON-state appropriately [31]. Figure 7 shows the voice activity model, while table 2 shows the parameter valuations.

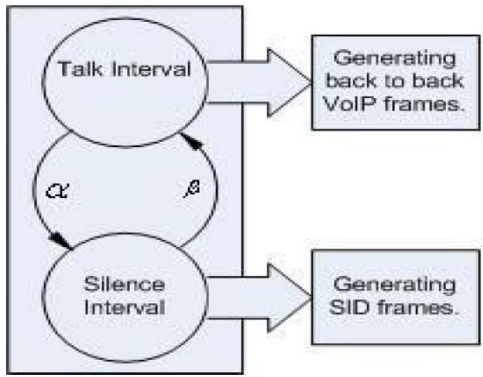

**Figure 7: Voice Activity Model [31]**

**Table 2: Exponential Voice Activity Model [31]**

| Parameters | Value |
|---|---|
| Mean Duration of ON-Period (1/α) | 352ms |
| Mean Duration of ON-Period (1/β) | 650ms |

With the adoption of ITU-T G.729A and B, numerous silence suppression algorithms have advanced with the objective reducing substantially the load on communication channel, by decreasing to a large extent (40-60%) the number of voice blocks generated by the codec [28], [32] thereby enhancing performance. Conversely, performance is not only the crucial issue, implementation complexity is a much an issue. The probable solution of which is; provisioning low-complexity silence suppression algorithms among other needs [30].

Consequently, appropriate features of VAD that makes better silence suppression schemes include; ***Good Decision Rule,*** A physical property of speech which when exploited yields stable and precise decision in categorizing fragments of the signal into silence or otherwise. ***Adaptability to Background Noise,*** this improves sturdiness, particularly in wireless mobile telephony. ***Low Computational Complexity,*** required for efficiency in real-time applications by [33] and [34].

One potential solution is presented in [33], comparing VAD algorithms for VoIP applications. The submission is based on efficient VAD scheme used for VoIP systems. The time domain VAD algorithms were seen to be computationally less composite, despite speech quality deterioration in comparison to frequency domain algorithms. The frequency domain algorithm is observed to sustain better immunity to low SNR compared to time domain algorithms. These outcomes informed the proposition of some VAD algorithms whose test outputs demonstrates the unfailing superiority of Comprehensive VAD scheme above all other schemes. Its lead over others is seen in better speech detection and noise immunity, although their still exist performance degradation under low SNR conditions, it is yet a better option for real-time applications [33].

### III. CONCLUSION AND FUTURE WORK.

Bandwidth utilization is indeed a key characteristic for improving quality of service in VoIP networks and applications; reasons being that communication operators and service providers continue to seek solutions that can lessen their odds at the same time attaining success-enabling positions in the current competitive environment. And given the varying techniques for improving bandwidth utilization (Voice / Packet Header Compression with aggregations, Reduction and Silence suppression) for which scalability, improved productivity, flexibility and reduction of service cost are key objectives. It is therefore significant that comparative approaches are able to yield maximum result with least delay and timely delivery of packets. Silence suppression is reasonably productive though leaves room for enhancement. Packet header reduction using multiplexing is seen to decrease both bandwidth wastage and the number of packets. The delta-multiplexing technique which combines the approach of header reduction and payload reduction yields a better result. However, the Packet (payload) compression technique produces the optimum result with codec relativities. The outcome even comes best when compression combines with aggregation technique. Hence it could be thought as the most proficient of the techniques examined. Probable improvements could tend towards the implementation of protocols, and codecs that would yield faster bit rates, reduced delays for compression and transformation.